# Bridging the gap between atomically thin semiconductors and metal leads

**Authors:** Xiangbin Cai[1]†, Zefei Wu[1]†, Xu Han[1,2], Shuigang Xu[1], Jiangxiazi Lin[1], Tianyi Han[1], Pingge He[1], Xuemeng Feng[1], Liheng An[1], Run Shi[1,3], Jingwei Wang[1,3], Zhehan Ying[1], Yuan Cai[1], Mengyuan Hua[4], Junwei Liu[1], Ding Pan[1,2], Chun Cheng[3], Ning Wang[1]*.

**Affiliations:**

[1]Department of Physics and Center for Quantum Materials, The Hong Kong University of Science and Technology, Clear Water Bay, Kowloon, Hong Kong, China.

[2]Department of Chemistry, The Hong Kong University of Science and Technology, Clear Water Bay, Kowloon, Hong Kong, China.

[3]Department of Materials Science and Engineering, Southern University of Science and Technology, Shenzhen 518055, China.

[4]Department of Electrical and Electronic Engineering, Southern University of Science and Technology, Shenzhen 518055, China.

*Correspondence to: phwang@ust.hk.

†These authors contributed equally to this work.

**Electrically interfacing atomically thin transition metal dichalcogenide semiconductors (TMDSCs) with metal leads is challenging because of undesired interface barriers, which have drastically constrained the electrical performance of TMDSC devices for exploring their unconventional physical properties and realizing potential electronic applications. Here we demonstrate a strategy to achieve nearly barrier-free electrical contacts with few-layer TMDSCs by engineering interfacial bonding distortion. The carrier-injection efficiency of such electrical junction is substantially increased with robust ohmic behaviors from room to cryogenic temperatures. The performance enhancements of TMDSC field-effect transistors are well reflected by the ultralow contact resistance (down to 90 $\Omega\mu m$ in $MoS_2$, towards the quantum limit), the ultrahigh field-effect mobility (up to 358,000 $cm^2V^{-1}s^{-1}$ in $WSe_2$) and the prominent transport characteristics at cryogenic temperatures. This method also offers new possibilities of the local manipulation of structures and electronic properties for TMDSC device design.**

With a geometry similar to that of graphene, atomically thin transition metal dichalcogenide semiconductors (TMDSCs) possess many valuable properties that further broaden the two-dimensional (2D) materials playground [1-3]. However, in contrast to graphene, the electrical performance of atomically thin TMDSC field-effect transistors (FETs), such as the carrier mobility at cryogenic temperatures, is generally insufficient to explore many exciting quantum transport properties or practical applications [4-6]. Electrically interfacing atomically thin TMDSCs with metal leads is inherently problematic because of undesired metal-semiconductor interface barriers. These include the tunnel and Schottky barriers as shown in Fig. 1 A and B, respectively. Fermi-level pinning effects and Schottky barriers naturally arise when defect-induced gap states occur at the interface [7,8]. These interfacial barriers dramatically suppress the carrier-injection efficiency by capping the available carrier mobility and presenting large contact resistance in TMDSC FETs.

There have been great efforts made over the past decade to minimize the electrical contact barriers in TMDSC FETs, which can be grouped into two major categories: (I) direct metallization of the contact region through doping treatments [9-11] or microscale phase transformation [12,13]. (II) van der Waals (vdW) contacts by inserting tunnel-barrier layers into the metal-semiconductor junction [14,15], or by using graphene/soft-landed metals as vdW-interfaced electrodes [16-22]. It is notable that through extra gating on the contact regions, graphene leads can eliminate Schottky barriers in few-layer $MoS_2$ FETs [16,17]. In this case, however, the inherent nature of vdW contacts still limits the charge carrier injection, because the large vdW gap (3-4 Å) between the metal lead and TMDSC surface acts as an additional tunneling barrier against the carrier flow.

An efficient carrier injection to the conduction or valence bands of TMDSCs requires effective orbital overlap or hybridization between the transition metal and the electrode atoms. For example, the conduction band minimum of monolayer TMDSCs arises mainly from the $d$-orbitals of transition-metal atoms [23], which are sandwiched by two layers of chalcogen atoms. Any carrier injection through the chalcogen surface of TMDSCs (i.e., across the vdW gap) is less efficient because the orbitals of chalcogen atoms contribute little to the energy band edges. Alternatively, an edge-contact strategy can offer the orbital hybridization advantage to transition-metal atoms [24]. However, Fermi-level pinning effects caused by the dangling bonds of TMDSCs edges inevitably result in severe Schottky barriers, degrading the FET performance at low temperatures (Fig. S1). Since rich quantum transport behaviors of fundamental importance, such as topological states and electron correlation effects, are completely perturbed by the thermal stimulation and the strong phonon scattering of atomically thin TMDSC channels at room temperature, probing fascinating condensed matter physics in TMDSCs requires the significant advance in their low-temperature contact performance.

To address these issues, we develop a local bonding distortion (LBD) strategy for realizing highly efficient electrical junctions with atomically thin TMDSCs down to the monolayer limit. Both atomic-resolution cross-section microscopy and Raman spectroscopy evidence the LBD as a nanoscale trigonal-prismatic-to-octahedral coordination change of metal-chalcogen polyhedra in the metal-TMDSC interface. The LBD region acts as a semi-metallic bridge between the metal lead and the pristine TMDSC channel, exhibiting robust ohmic behaviors from room to cryogenic temperatures. The LBD contacts significantly elevate the carrier-injection efficiency of TMDSC FETs, demonstrating the ultralow contact resistance (down to 90 $\Omega\mu$m in 3L-MoS$_2$), the ultrahigh field-effect mobility (up to 358,000 cm$^2$V$^{-1}$s$^{-1}$ in 5L-WSe$_2$) and prominent transport characteristics at 0.3 K. This work not only paves the way to exploring the unconventional quantum transport properties in TMDSCs, but also provides a new scheme in the local manipulation of structure and electronic properties of TMDSCs.

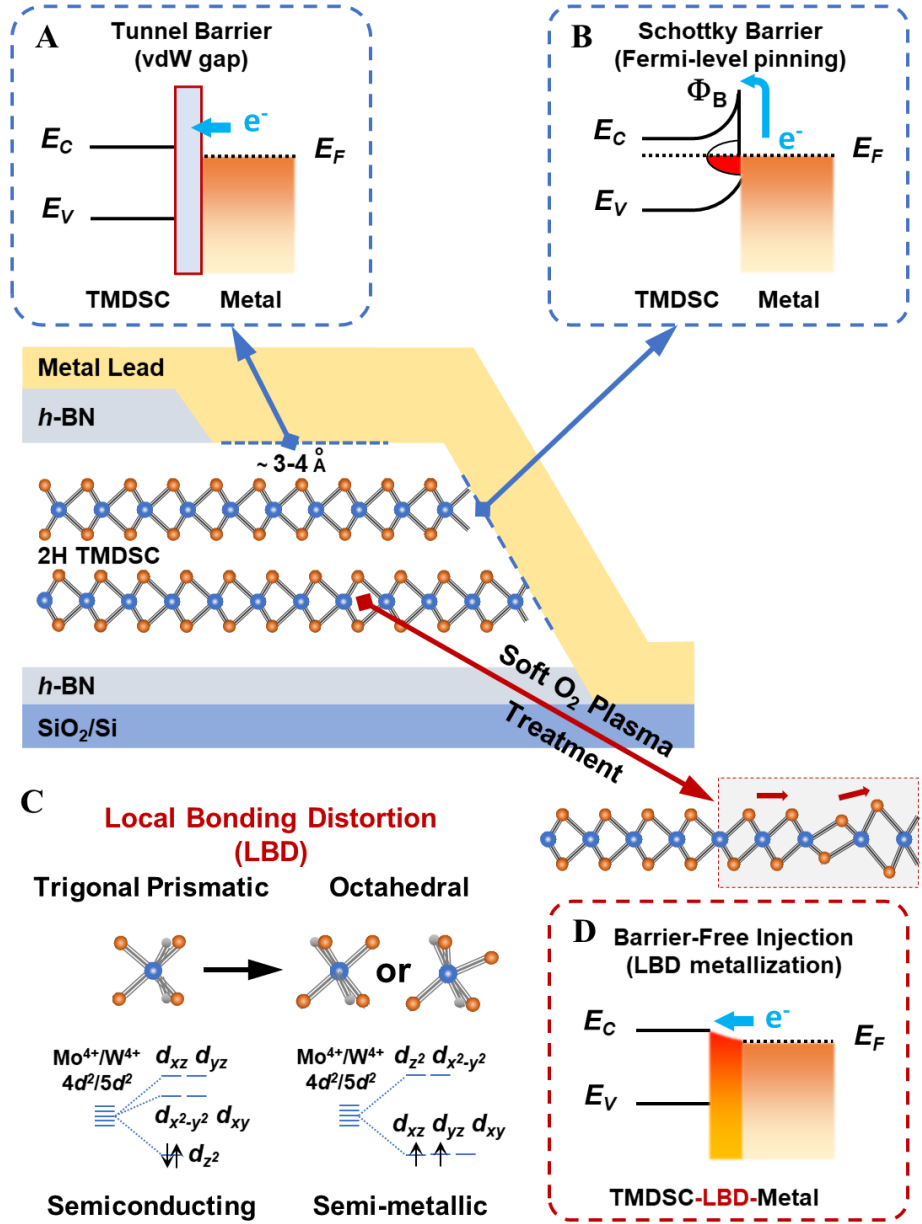

**Fig. 1: Different electrical interfaces between atomically thin TMDSCs and metal leads.** (*A*) *Interfacial band alignment of the van der Waals (vdW) contact.* (*B*) *Interfacial band alignment of the Schottky-limited contact.* (*C*) *Atomic configurations of [TM-C₆] polyhedra and corresponding energy splitting diagrams of Mo/W d-orbitals before and after the local bonding distortion (LBD).* (*D*) *Interfacial band alignment of the contact utilizing LBD mechanism.*

First of all, we would like to introduce the basic mechanism of our LBD strategy. As shown in Fig. 1 C, due to the trigonal-prismatic coordination of [TM-C$_6$] polyhedra in the H structure (in which TM stands for the transition metals, and C stands for the chalcogen atoms), the TM$^{4+}$ $d$-orbitals exhibit a close-shell electronic configuration (two $d$-electrons with opposite spins in one orbital) without itinerant electrons, resulting in semiconducting properties. In contrast, the octahedral coordination after distortion owns a rearranged energy splitting for the TM$^{4+}$ $d$-orbitals. The partial filling of two $d$-electrons with parallel spin into the three-fold degenerate orbitals provides both unoccupied states and itinerant electrons that increase the metallicity of the material [27]. The local distortion of [TM-C$_6$] polyhedra from trigonal-prismatic coordination to octahedral one modulates the electronic property into semi-metallicity, offering dispersed density of states (DOS) along the TMDSC energy band gap, and thus forms nearly barrier-free electrical interfaces as illustrated in the interfacial band diagram of Fig. 1D. Such lattice distortion is different from destructive defects and has been applied to engineer bulk materials' properties [25,26], while it has not been reported in TMDSCs yet. It is also different from the common H-to-T phase transition in TMDSCs by its nanoscale localization.

Technically, we chose the oxygen (a chalcogen element) plasma to trigger the LBD in contact interfaces of several representative 2D semiconductors, including MoS$_2$ and WSe$_2$. Schematics in Fig. 2 show the work flow of our referenced RIE processes to fabricate LBD in both edge- and top-contact geometries. A thin poly(methyl methacrylate) (PMMA) mask is defined by e-beam lithography for the simultaneous reactive ion etching (RIE) of contact windows and the reference area. By monitoring the optical contrast of the reference area, either a top-BN/TMDSC/substrate region in the edge-contact geometry or a top-BN/substrate region in the top-contact geometry, the etching depth in contact windows is controlled precisely (the optical contrast of 300-nm-SiO$_2$/Si substrate will be totally different without 2D materials coverage). Followed by the tuned soft oxygen-plasma treatment, LBD appears on the exposed TMDSC edges or surface for different contact geometries. We want to emphasize that the oxygen-plasma treatment is configured in a soft-landing manner to avoid generating defects in TMDSCs by paralleling the capacitively coupled electrostatic field along the sample surface, reducing the plate bias and lowering the input power. More details of the FET fabrication can be found in the Method section and step-by-step images in Fig. S2. The LBD is reproducible once suitable conditions are set up. With reasonable proposals, we are pleased to provide devices utilizing LBD contacts for collaborative research.

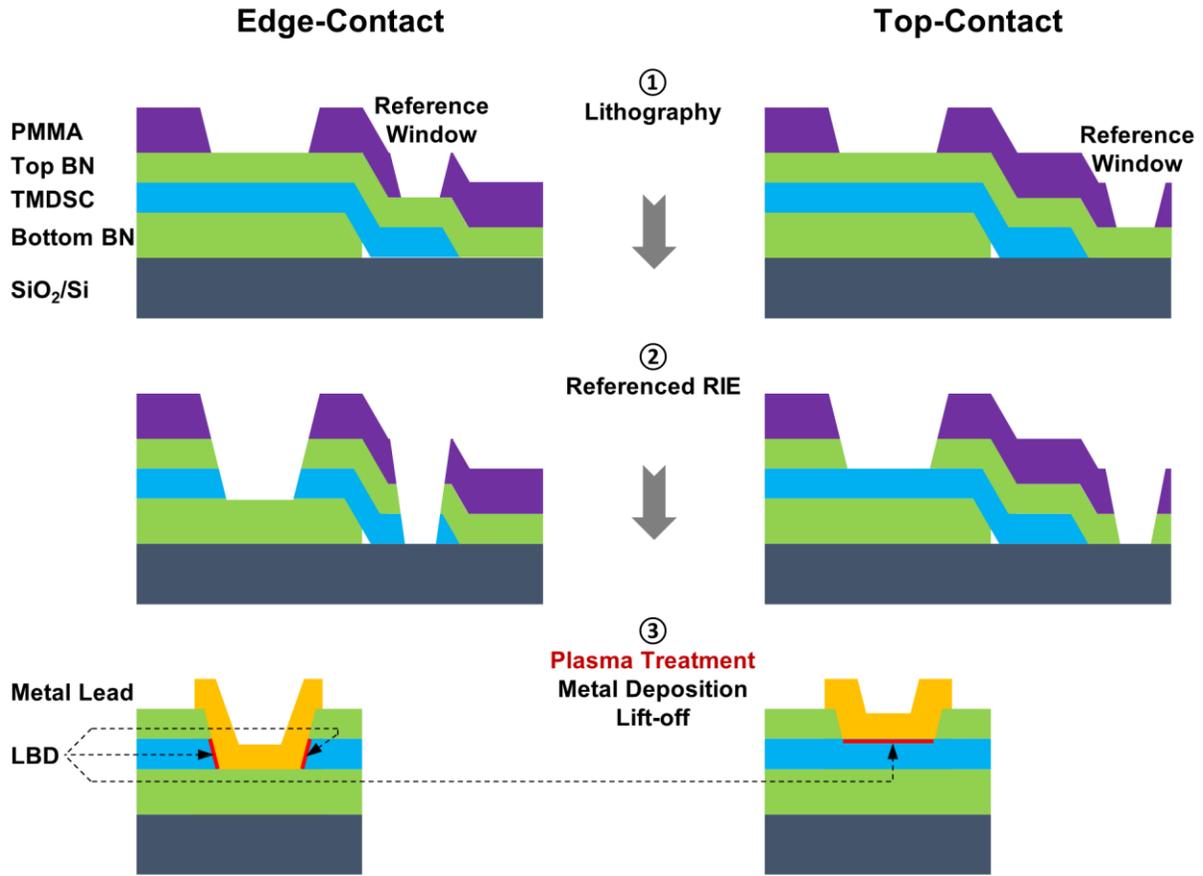

**Fig. 2: Mechanism of the referenced RIE technique.** *Schematics showing the work flow of fabricating LBD contacts to BN-encapsulated TMDSCs in both edge- and top-contact geometries. After the soft oxygen-plasma treatment, LBD appears on the exposed TMDSC edges or surface for the edge- and top-contact geometries, respectively.*

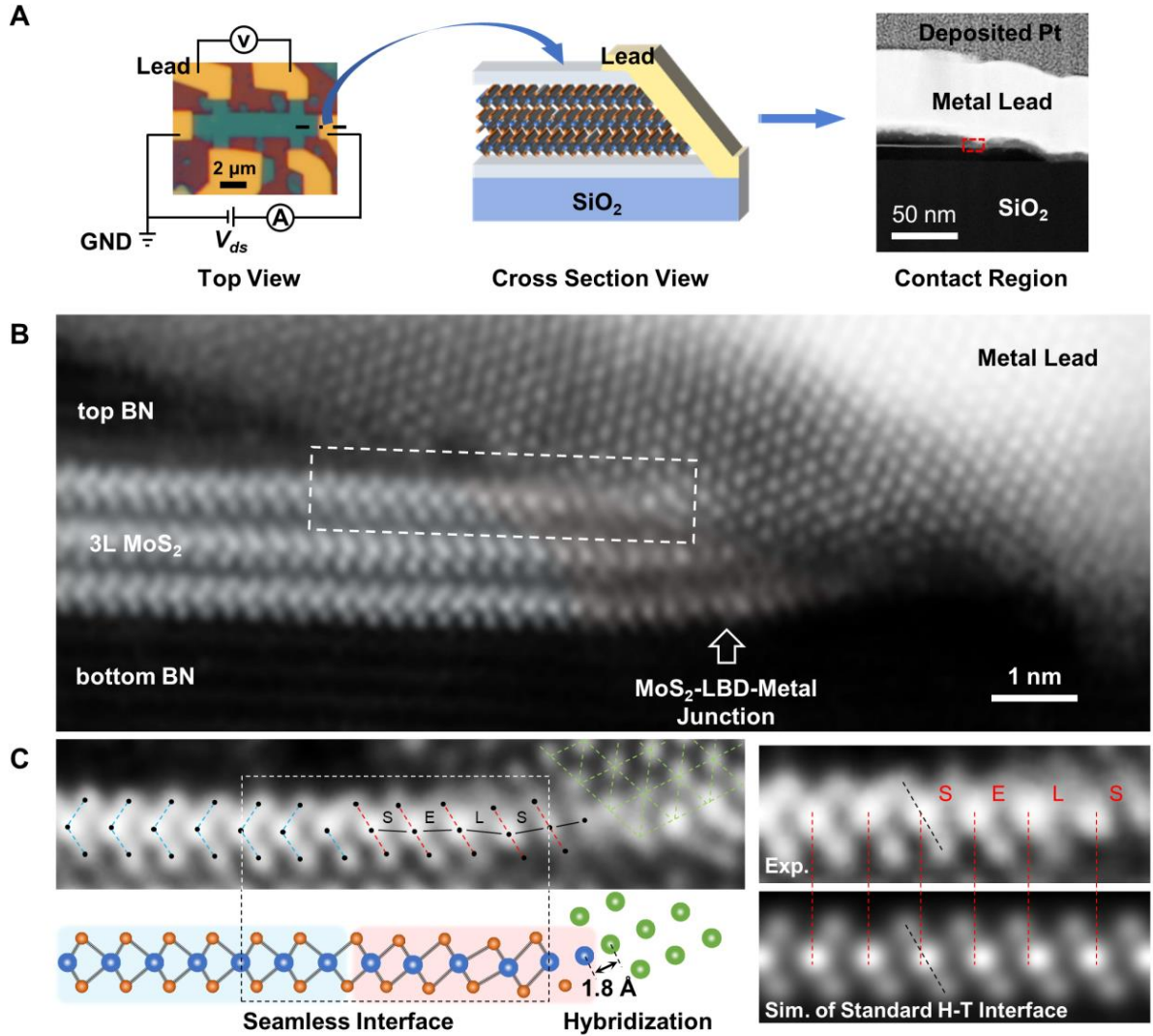

**Fig. 3: Atomic-scale observation of the LBD contact in MoS₂.** (*A*) *Top-view optical image of a typical 3L-MoS$_2$ FET. The dash-dot line indicates the cutting position in the source, from which a cross-section membrane (as illustrated in the middle schematic) was extracted using the focused ion beam (FIB) method. The low-magnification ADF image of the contact region is shown on the right. The area in the red-dash rectangle is zoomed in below. (**B**) Atomic-resolution ADF image of the MoS$_2$-LBD-metal junction observed along the MoS$_2$ zig-zag direction. The first layer of MoS$_2$ is analyzed in (**C**) with the corresponding atomic model shown below. The Mo-Mo bond-length comparison (E: equal, S: short, L: long) between the experimental measurement and the simulated standard structure is shown on the right.*

Our LBD contact strategy works perfectly in the edge-contact geometry for exploiting its strong orbital overlap and hybridization advantages as demonstrated by 3L-MoS$_2$ FETs. In the top and cross-section views of Fig. 3 (A), after electrical measurements, a cross-section membrane from one of the main electrical contacts was lifted out for the cross-section observation. In the low-magnification annual dark field (ADF) image of the contact region, the elegant edge-cutting of the BN-MoS$_2$-BN stack by the referenced RIE process can be observed. The electrical interface between the 3L-MoS$_2$ edge and the deposited metal was observed at atomic resolution along the zig-zag direction, as shown in Fig. 3 (B). The atomic-resolution ADF image shows three main structural features of the plasma-induced LBD in the edge-contact region: (1) the LBD localizes within the ~1 nm wide edge of 3L-MoS$_2$; (2) both Mo and S atomic positions deviate from the ideal 2H phase on the left and form octahedral configurations (close to T phase); (3) terminated Mo atoms at the MoS$_2$ edge connect directly to the lead surface with a closest distance of ~1.8 Å, which is much smaller than the normal vdW gap size (~3-4 Å) and similar to that of a typical chemical bond. This atomic-scale observation supports the strong orbital overlap and hybridization between transition metal and electrode atoms in an edge-contact geometry.

The first layer of MoS$_2$ was further analyzed in Fig. 3 (C) in comparison to the simulated standard H-T interface. Since the atomic structure on the left H region remains identical with the simulated one, there should be no damage or artefact introduced by our sample preparation processes. Both H-T hetero-phase interfaces are structurally seamless without dangling bonds or defects, where only slight stretching or shrinking of some bond lengths is involved in the distortion. This 1-nm-wide structural change looks similar to the 1T phase transition, but the Mo-Mo bond angle, length and interlayer stacking order differ from those of the 1T phase as compared with the simulated standard H-T interface. In specific, the Mo-Mo spacings in the experimental measurement stagger around the equal value in an ideal 1T phase. The first unit cell in the termination shrinks a bit while the second one on the left elongates. The slight deviation of atomic positions from standard crystallographic sites as observed here should affect the electronic property through the electron-lattice coupling, which makes the DOS of such semi-metallic nanostructure dispersed away the theoretical work-function value of ideal crystal. The energy dispersion of DOS thus allows the nearly barrier-free electron transport from the metal Fermi level to the TMDSC conduction band edge as illustrated in the interfacial band diagram of Fig. 1 (D).

By performing the electron energy-loss spectroscopy (EELS) analysis in this LBD contact region as shown in Fig. S3, we found there is some oxygen doping into the distorted $MoS_2$ edge, which is consistent with the proposed formation mechanisms of LBD by the oxygen-substitution-induced bond rearrangements.[28] That means the LBD may come from the spontaneous bond rearrangement of TMDSCs after the partial substitution of chalcogen atoms by oxygen during the soft $O_2$-plasma treatment. As depicted in Fig. S4A, only stretching or shrinking of some bond lengths and angles is involved in the LBD, which in turn balances the size difference between chalcogen and oxygen atoms after the substitution. The $O_2$-plasma-induced LBD was further confirmed by micro Raman spectroscopy (Fig. S4B). The local distortion of [$TM$-$C_6$] polyhedra from the trigonal-prismatic coordination to octahedral ones modulates the electronic property into semi-metallicity, boosting the carrier-injection efficiency of electrical contacts and thus device performance as discussed below.

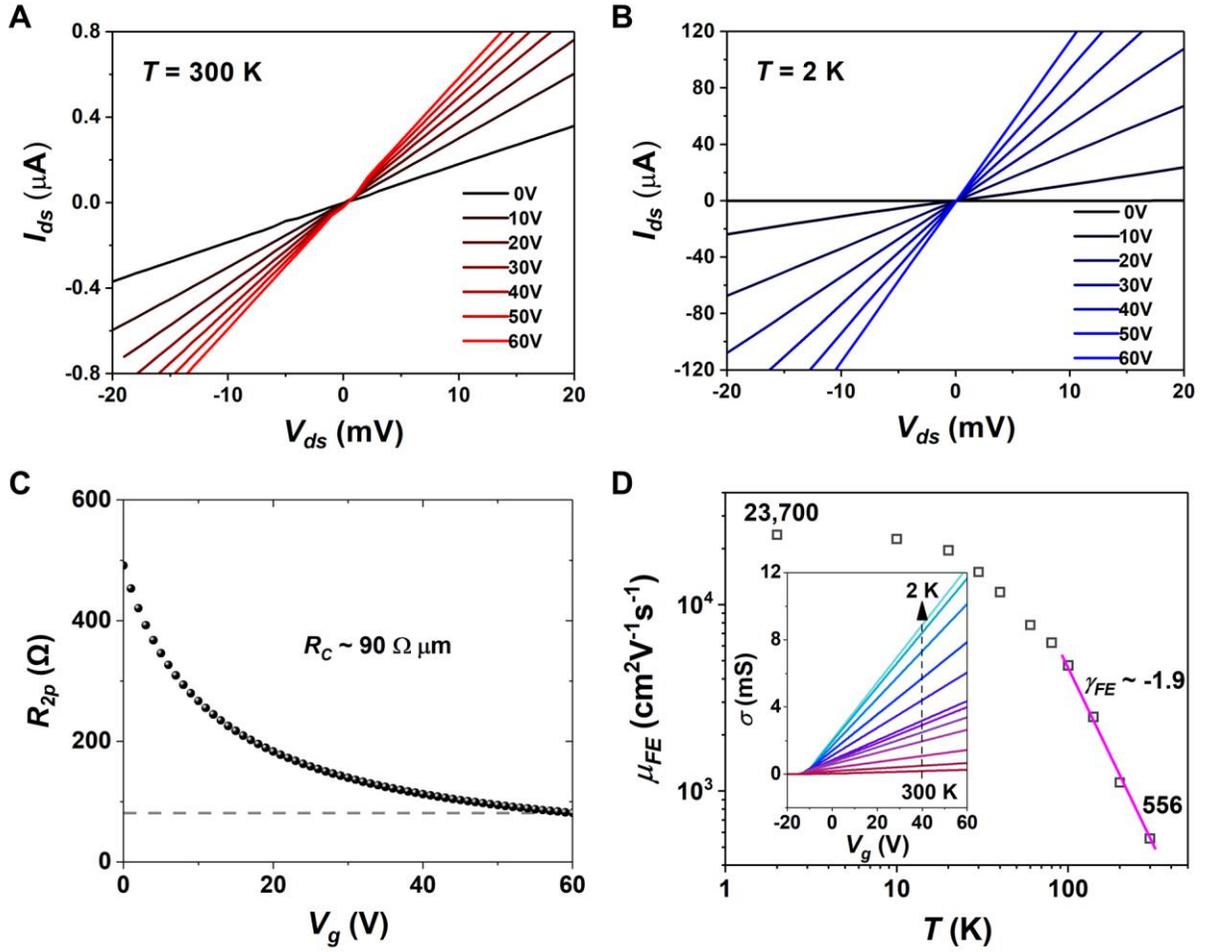

**Fig. 4: Electrical performance of the 3L-MoS$_2$ FET utilizing LBD contacts.** *Two-probe I$_{ds}$-V$_{ds}$ curves at (**A**) room temperature and (**B**) 2 K, respectively. (**C**) Two-probe resistance, R$_{2p}$, as the function of gate voltages. The contact resistance can be estimated from the highly gating regime, where the channel is made conductive by the heavy electrostatic doping and its resistance should be negligible. (**D**) Temperature-dependent field-effect mobility, μ$_{FE}$, derived from the channel-conductance as the inset shows, indicating the strong phonon scattering effects of 3L-MoS$_2$ at high temperatures.*

The 3L-MoS$_2$ FET interfaced by the LBD contacts shows excellent performance at both room and cryogenic temperatures. In Fig. 4 (A) and (B), linear two-probe output curves, $I_{ds}$-$V_{ds}$, can be observed at all gate voltages and well maintained down to 2 K, demonstrating the robust ohmic contact nature. The channel current at 2 K is about 200 times larger than that at 300 K, which suggests little contribution from the contacts to the total device resistance. In Fig. 4 (C), the two-probe resistance (including both channel and contact resistance) at 2 K decreases to around 90 Ω as the back-gate voltage increases, indicating the upper bound of the contact resistance of around 90 Ωμm, given the negligible channel resistance under the heavy electrostatic doping. We notice that the quantum limit of the contact resistance is inversely related to the 2D charge carrier density ($n$), yielding $0.026/\sqrt{n} \approx 30$ Ωμm when $n = 10^{13}$ cm$^{-2}$, which suggests that our LBD contact quality is approaching the theoretical limit.[4] The channel conductance in Fig. 4 (D) inset increases steadily with the lowered temperature, reflecting the gradually lifted phonon scattering in the MoS$_2$ channel. The extracted field-effect mobility shows ultrahigh values of 556 cm$^2$V$^{-1}$s$^{-1}$ at room temperature and 23,700 cm$^2$V$^{-1}$s$^{-1}$ at 2 K with a phonon-scattering constant of around -1.9, confirming the strong phonon-electron interaction in atomically thin MoS$_2$.

This contact scheme also works well for monolayer MoS$_2$ as demonstrated in Fig. S5. The metal-LBD-MoS$_2$ interfaces effectively lift the carrier-injection barriers in 1L-MoS$_2$ FETs, manifesting high-quality quantum-Hall detection, brilliant subthreshold swings, ultrahigh field-effect and Hall mobilities (9,900 and 9200 cm$^2$V$^{-1}$s$^{-1}$, respectively) at cryogenic temperatures. Such high-performance 1L-MoS$_2$ devices by the LBD method should facilitate future study on the electron-electron interaction effects at fractional quantum Hall states. The relatively suppressed carrier mobility for 1L-MoS$_2$ may be due to the weaker screening effect of monolayer channels and thus increased scattering by intrinsic impurities, but the effectiveness of LBD strategy for monolayer TMDSCs is verified.

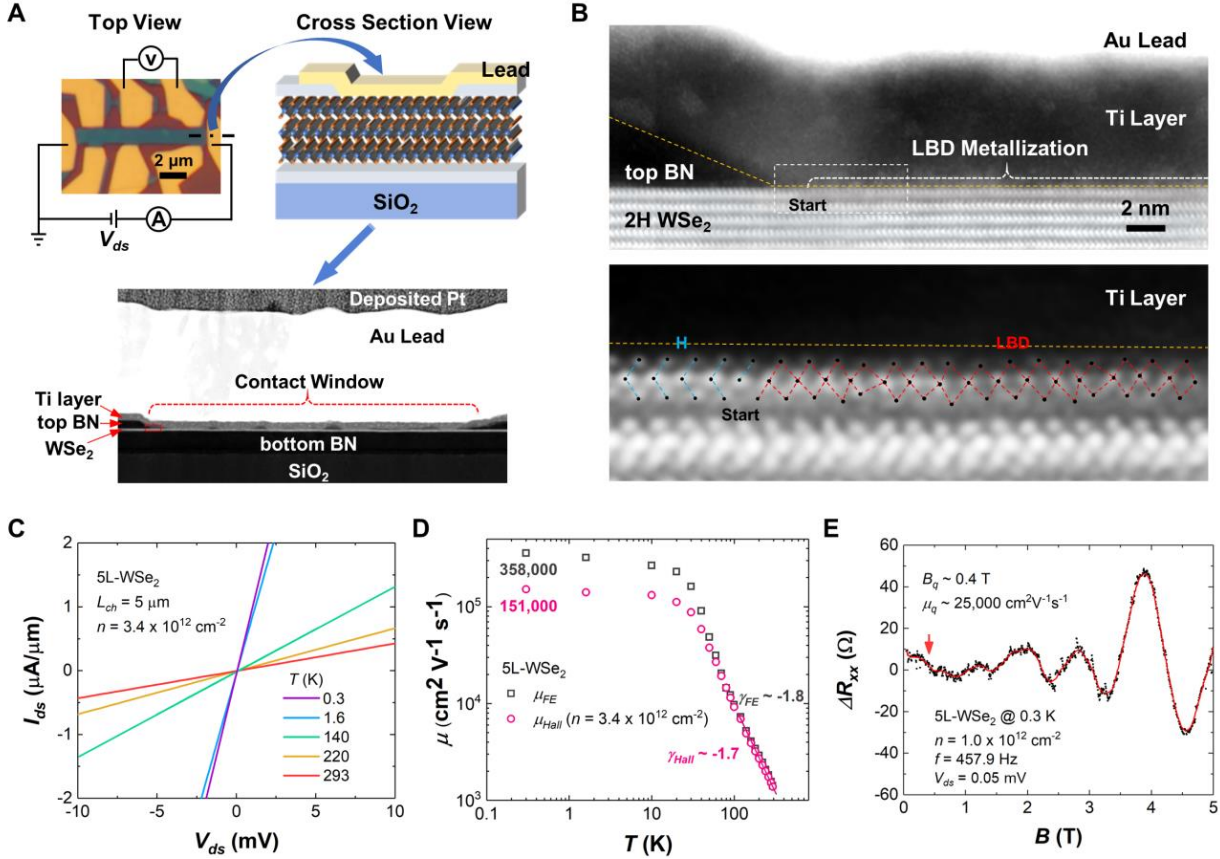

***Fig. 5: Generality of the LBD strategy.*** (***A***) *Optical image of a typical 5L-WSe₂ FET utilizing LBD contacts. The dash-dot line indicates the cutting position in the source, from which a cross-section membrane was lifted out as illustrated on the right. The low-magnification ADF image of the contact region is shown below. The area in the red-dash rectangle is zoomed in on the right.* (***B***) *Atomic-resolution ADF image of the WSe₂-LBD-metal junction observed along the WSe₂ zig-zag direction. The orange dashed line shows the bottom of Ti layer. The LBD-starting region is analyzed below with the contrast re-surveyed, and the atomic model overlaid to clarify the distorted bonding.* (***C***) *Tow-probe I$_{ds}$-V$_{ds}$ curves at varied temperatures.* (***D***) *Log-scale plots of field-effect and Hall mobilities as the function of temperatures.* (***E***) *Shubnikov de Haas (SdH) oscillations of the longitudinal channel resistance at low magnetic field strength down to ~0.4 T, indicating the ultrahigh quantum mobility.*

Furthermore, we demonstrate the generality of LBD strategy using 5L-WSe$_2$ in the more common top-contact geometry. The top-contact geometry is widely used as it provides a much larger carrier-injection area than the edge-contact scheme. The precise removal of top BN to expose the pristine WSe$_2$ surface is realized by the referenced RIE process as illustrated in Fig. 2. From the low-magnification ADF image of the contact region in Fig. 5 (A), a uniform top-contact area can be identified, except for some top BN islands due to nanoscale thickness differences. After the soft oxygen-plasma treatment, the topmost WSe$_2$ layer within the contact window went through the LBD without affecting the layers beneath as the atomic-resolution ADF images of metal-WSe$_2$ electrical junction in Fig. 5 (B) show. This LBD metallization exhibits complicated changes of the W-Se bonding without any defects (e.g. vacancies or dangling bonds). According to the 2D-Gaussian-fitted W and Se atomic column positions overlaid in the image (see Fig. S6 for the image-processing details), the newly formed structure turns out to be a mixture of 1T and 1T' octahedral derivatives. We notice that the nanoscale T/T'-mixed structure does not create defects and thus can be considered as a distorted crystalline structure. Such distorted WSe$_2$ layer acts as a semi-metallic bridge between the metal lead and the pristine WSe$_2$ channel, leading to the nearly barrier-free carrier-injection, similar to the cases in MoS$_2$.

Linear output characteristics are observed across the wide temperature range as shown in Fig. 5 (C), proving the robust ohmic contact nature. Although the contact resistance of 700 $\Omega\mu$m is larger than that in edge-contacted MoS$_2$ due to the inevitable vdW gap in the top-contact geometry (see Fig. S7 for extended data from this top-contacted WSe$_2$ FET), we are amazed that ultrahigh field-effect ($\mu_{FE}$ ~358,000 cm$^2$V$^{-1}$s$^{-1}$), Hall ($\mu_{Hall}$ ~151,000 cm$^2$V$^{-1}$s$^{-1}$) and quantum ($\mu_q$ ~25,000 cm$^2$V$^{-1}$s$^{-1}$) carrier mobilities are achieved at 0.3 K, as the temperature plots in Fig. 5 (D) and the prominent Shubnikov de Haas (SdH) oscillations in Fig. 5 (E) demonstrate. Although the quantum mobility is a rough estimation from the oscillation onset at low magnetic field, the derived value should be the lower bound, because the residual electrical turbulence from both the measurement system and the external environment will bury the real onset of oscillation amplitude smaller than the noise level. Theoretically, there is $\mu_{FE} > \mu_{Hall} > \mu_q$ for the same sample and temperature. The field-effect mobility is always higher than the Hall mobility of the same conditions because of $\mu_{FE}$ = $\mu_{Hall} + n \frac{d\mu_{Hall}}{dn}$, where higher carrier density can screen the impurity scattering and results in $\frac{d\mu_{Hall}}{dn} > 0$. On the other hand, since the quantum scattering time is always shorter than the transport scattering time, quantum mobility should fall below the corresponding Hall mobility. No hysteresis effects were observed when scanning gate voltage forwards and backwards as shown in Fig. S7 (C), which confirm there are no chargeable states caused by the LBD, in consistence with the atomic observation of no dangling bonds or vacancies in the metal-TMDSC junction. Furthermore, the reproducible contact quality after long-time storage as shown in Fig. S7 (D) and the competitive performance of multiple devices with the same contact geometry and different channel thicknesses shown in Fig. S8 demonstrate the excellent reproducibility and practicality of our LBD contact strategy for TMDSCs.

The high performance of the electrical interfaces formed by the LBD method is attributed to the octahedral distortion of TMDSCs. Unlike a perfect H-T interface, in which the work functions of bulk 1T and H phases are mismatched, various sizes of distorted T/T' derivatives induced by the oxygen plasma are semi-metallic, offering dispersed energy states around its work function. These energy states may play an important role in the orbital hybridization between transition metals and leads atoms, and thus enhancing the carrier-injection efficiency. This mechanism is supported by the fact that the polarity of the LBD interfaced FETs is contact-metal dependent. For example, Ti/Au leads can access the conduction band edge (n-type) while Pd leads can access the valence band edge (p-type) of $WSe_2$ (Fig. S9). The DOS of the LBD region is distributed along the energy band gap of TMDSCs, easily coupled with different work functions of metal leads. We also measured the resistance of $1L-MoS_2$ channel after treated by the same soft oxygen-plasma process as fabricating LBD contacts at varied temperatures (Fig. S10), where the LBD-induced metallicity and the strong electron-phonon coupling are shown. For the standard bulk T-H interface, however, the device polarity is independent on contact metals used [12] since the work function of 1T phase is fixed, resulting in limited accessibility to the energy bands of TMDSCs.

In conclusion, we demonstrate a proof-of-concept LBD contact strategy by the soft oxygen-plasma treatment, which is flexible for both edge- and top-contact fabrication schemes, to construct high-quality electrical junctions between metal leads and different TMDSCs, down to their monolayer limit. The excellent performance of such electrical contacts is structurally understood by the cross-section observation of metal-semiconductor junctions from practical devices. By exploiting such high-quality contacts, we have achieved robust ohmic behaviors, the ultralow contact resistance (down to 90 $\Omega\mu m$ in $3L-MoS_2$), the ultrahigh mobilities (up to 358,000 $cm^2V^{-1}s^{-1}$ in $5L-WSe_2$) and prominent transport characteristics. Since the electrical contact quality at cryogenic temperatures has been considered as the main factor constraining the study of unconventional quantum transport properties in TMDSCs, our LBD contact strategy should facilitate the 2D-material physics study. Importantly, this method is fully compatible with the clean-room processes for future scalable electrode integration and may shed light on novel vdW device design.

## Methods:

### Materials

TMDSC crystals were bought from the website www.2dsemiconductors.com while the hexagonal BN (grade A1) was bought from the website www.hqgraphene.com.

### FET Fabrication

The sandwich structure of top BN (5-8 nm), few-layer 2H-MoS$_2$/WSe$_2$ and bottom BN (12-20 nm) was assembled by the dry pickup-transfer technique in a glove box. Following the electron-beam lithography patterning of a thin poly(methyl methacrylate) (PMMA, A5) mask using Raith eLiNE, the heterostructure was first shaped into Hall-bar structure by the reactive ion etching (RIE) in STS Pro using the gas mixture of CHF$_3$ and O$_2$ (40:4 sccm). Repatterning a PMMA mask then defined the contact windows and a reference area for the referenced RIE as illustrated in Fig. 2. Before the deposition of metal leads, the TMDSCs were exposed to the pure oxygen-plasma flow for around 10 seconds in a soft-landing manner, which was realized by paralleling the electrostatic field along the sample surface, reducing the plate bias and lowering the input power. The non-destructive high-vacuum metal deposition [19] was performed by Peva 450E right after the oxygen-plasma treatment to seal the contact windows.

### Electrical Measurements

The DC $I_{ds}$-$V_{ds}$ curves were measured in the two-probe configuration using Keithley 6430 while the channel conductance ($\sigma$) and magneto-transport measurements were conducted in the four-probe configuration using lock-in techniques (Stanford Research 830 as the amplifier and DS 360 as the function generator). The cryogenic station from Oxford Instruments was used to provide cryogenic temperatures down to 1.4 K and magnetic fields up to 15 T. A home-modified He3 holder provides stable 0.3 K environment for ~24 h per regeneration.

Different types of carrier mobilities were calculated to assess the contact performance, including field-effect, Hall and quantum mobilities. The field-effect mobility ($\mu_{FE}$) values were obtained according to $\mu_{FE} = \frac{1}{C}\frac{\Delta\sigma}{\Delta V_g}$, where $C$ is the gate capacitance (from serially-connected capacitors of the 300 nm SiO$_2$ and the 12-20 nm bottom BN layer for back-gated devices), which was accurately determined by the Hall effect measurements of carrier density versus gate voltage relationship. The $\frac{\Delta\sigma}{\Delta V_g}$ is the slope of channel conductance ($\sigma = \frac{GL}{W}$, where $G$ is the four-probe conductance, $L$ is the channel length and $W$ is the channel width of inner probes) versus gate voltage at the linear region. The Hall mobility ($\mu_{Hall}$) values were obtained according to $\mu_{Hall} = \frac{\sigma}{ne}$, where $n$ is the carrier density derived from the Hall coefficient, $e$ is the elementary charge. The quantum mobility ($\mu_q$) value was derived from the lowest perpendicular magnetic field strength $B_q$ where SdH oscillations onset, as $\mu_q = \frac{1}{B_q}$. The exponential temperature decay of both field-effect and Hall mobilities follows $T^\gamma$, where $\gamma$ indicates the phonon scattering strength.

The contact resistance was extracted from the two-probe resistivity or the resistance difference between two-probe and four-probe measurements according to 2*$R_c$ = ($R_{2p}$ − $\frac{L_{out}}{L_{in}}R_{4p})*W$, where $R_{2p}$ and $R_{4p}$ are the two-probe and four-probe resistance, respectively; $L_{out}$ and $L_{in}$ are the distance of outer probe pair and inner probe pair in the four-probe configuration, respectively.

**Cross-section Preparation**

Focused ion beam (FIB) techniques using FEI Helios G4 UX were applied to prepare the cross-sectional membranes from FET contact sites, after electrical measurements. Before ion milling processes, a low-energy (2 kV) electron beam was used to identify the exact contact positions and deposit 500-nm-thick Pt protecting straps on the top of metal leads. Then thicker Pt protecting straps (1.5 μm thick) were deposited using the ion beam. After lifting out a 1-μm-thick plate from the chosen contact site to a copper finger, step-by-step thinning processes using 30, 5, 2, 1 and 0.5 kV ion beams well minimized the beam damage to the 2D-material heterostructures. Finally, an ex-situ 500 V argon-ion beam shower of 30 s was used to further clean both membrane surfaces. According to the low-loss electron energy-loss spectroscopy (EELS), typical sample thickness at the contact interface is 12.3 ± 2.5 nm, which guarantees atomic-resolution imaging and elemental mapping at 60 kV acceleration voltage.

**STEM Characterization**

JEOL JEM ARM 200F (ACCELARM) equipped with a cold field-emission gun and the ASCOR fifth-order probe corrector was utilized for the aberration-corrected scanning transmission electron microscopy (ACSTEM) study of the contact interface at atomic resolution under 60 kV acceleration voltage. Imaging and spectroscopy were performed using 32 mrad convergence semi angle and 24-60 pA probe current for the optimal information transfer and minimal electron irradiation. The low acceleration voltage of 60 kV and electron dose of ~7.5-20.0 Me nm$^{-2}$ are far away from the threshold to cause electron-induced effects at room temperature. [29] The collection semi angles for annular dark field (ADF) imaging were set to 68-200 mrad and the Gatan Enfinium spectrometer was used for the acquisition of energy-loss spectra under the dual EELS mode.

**Image Analysis**

The atomic positions in model drawings of Fig. 3 (C) and Fig. 5 (B) were extracted by the 2D-gaussian fitting of normalized intensities of atomic columns in the HAADF images. We show the data-processing workflow and distortion-analysis examples in Fig. S6. The method can provide picometer precision for the fitted atomic positions when using an aberration-corrected STEM. Based on the extracted atomic positions, the LBD of oxygen-plasma-treated $MoS_2$ and $WSe_2$ can be identified by bond length/angle variations unit-cell by unit-cell in comparison to the simulation of standard structures.

**Image Simulation**

Multislice codes by Earl J. Kirkland were used for the simulation of atomic-resolution ADF images to handle the dynamical scattering processes between matter and transmitted electrons. [30] 2H, 1T and 1T' structure files were downloaded from the American Mineralogist Crystal Structure Database. An electron probe of experimental conditions was focused on the crystal surface and the scattered electrons within 68-200 mrad collection semi angles were integrated. Twenty frozen phonon configurations were applied to represent the thermal diffuse scattering effects at room temperature. The effective source size effect was included by convolving simulated images with a 2D Gaussian function of 0.8 Å FWHM.

**DFT Calculations**

The plane-wave pseudopotential method implemented in the Quantum ESPRESSO package (version 6.1) [31] was used to calculate electronic structures. The SG15 Optimized Norm-Conserving Vanderbilt (ONCV) pseudopotentials were used here. [32,33] Full relativistic pseudopotentials were employed in the calculations with spin-orbital coupling. [34] The kinetic energy cutoff for plane waves was 60 Ry. The convergence thresholds for energy, force and stress were $10^{-5}$ Ry, $10^{-4}$ Ry/Bohr and 50 MPa, respectively. The multilayer structures were obtained from full relaxation with the van der Waals functional optB88. [35] PBE exchange-correlation functional was adopted to calculate electronic structures. [36]

**Acknowledgments:**

The work is financially supported by the Hong Kong Research Grants Council (Projects No. AoE/P-701/20, 16303720 and C6025-19G), the National Key R&D Program of China (2020YFA 0309600) and the William Mong Institute of Nano Science and Technology. C.C. thank the funding support from the National Natural Science Foundation of China (Grants No. 91963129 and 51776094).

**Author Contributions:**

N.W. conceived and directed the project. Z.W., X.C., J.L. and T.H. conducted the device fabrication and electrical measurements. X.C. performed the FIB sample preparation and ACSTEM characterization. X.H., J.L. and D.P. carried out the DFT calculations. R.S. grew the $MoS_2$ monolayer crystals. X.C. and N.W. wrote the paper. All authors participated in the data analysis, result discussion and manuscript preparation.

**Additional Information:**

Supplementary materials, including the text on more results discussion and the supporting figures (Fig. S1-S10), are available. The authors declare no competing financial interests. Correspondence and request for materials should be addressed to N.W. at phwang@ust.hk.

Supplementary Materials for

**Bridging the gap between atomically thin semiconductors and metal leads**

**This PDF file includes:**

- ❖ Fig. S1: Control device of 3L-MoS$_2$ by direct edge contacts
- ❖ Fig. S2: Sample fabrication steps
- ❖ Fig. S3: EELS analysis of the LBD contact region
- ❖ Fig. S4: Raman spectroscopy of the LBD
- ❖ Fig. S5: 1L-MoS$_2$ FET by LBD contacts
- ❖ Fig. S6: ACSTEM image analysis
- ❖ Fig. S7: Extended data of the 5L-WSe$_2$ FET using LBD contacts
- ❖ Fig. S8: Reproducibility and practicality of LBD contacts in TMDSCs
- ❖ Fig. S9: Lead-dependent polarity of WSe$_2$ FETs
- ❖ Fig. S10: LBD-induced metallicity in TMDSC monolayers

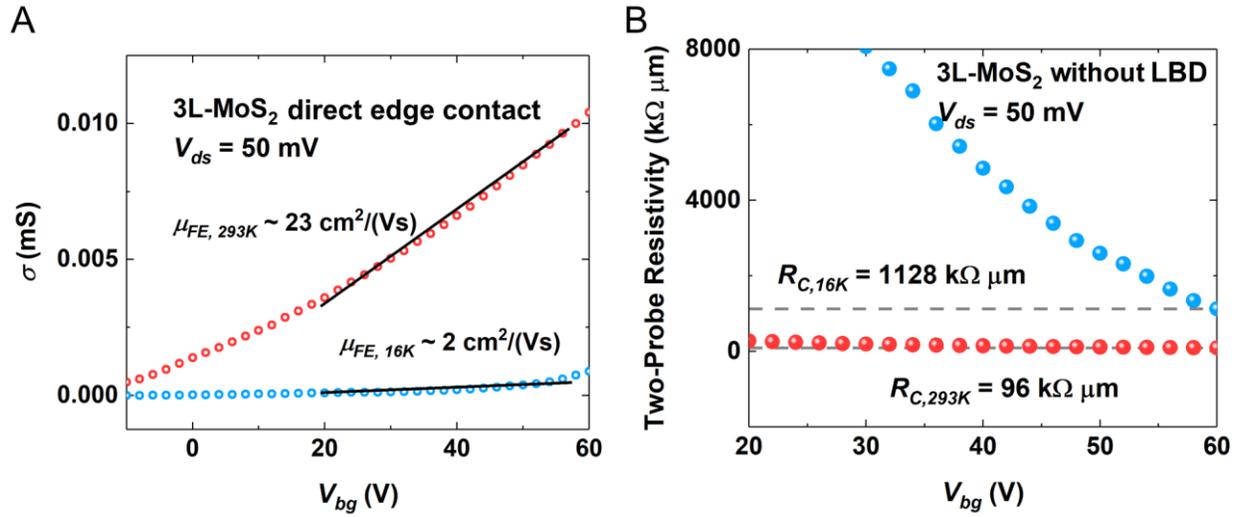

**Fig. S1: Control device of 3L-MoS₂ by direct edge contacts.** (**A**) Field-effect mobility and (**B**) contact resistance at room and cryogenic temperatures. The direct electrical interfacing between metal leads and semiconducting MoS₂ edges causes severe Fermi-level pinning and Schottky barriers, as the band diagram in Fig. 1 B shows, which degrade the FET performance at cryogenic temperatures. As shown in Fig. S1, the channel conductance was low with four-order smaller mobilities and larger contact resistance than those utilizing LBD contacts reported in Fig. 4. The device performance becomes worse at lower temperatures as the main carrier-injection mechanism in the contact interface shifts from the thermionic emission to the tunneling across Schottky barriers.

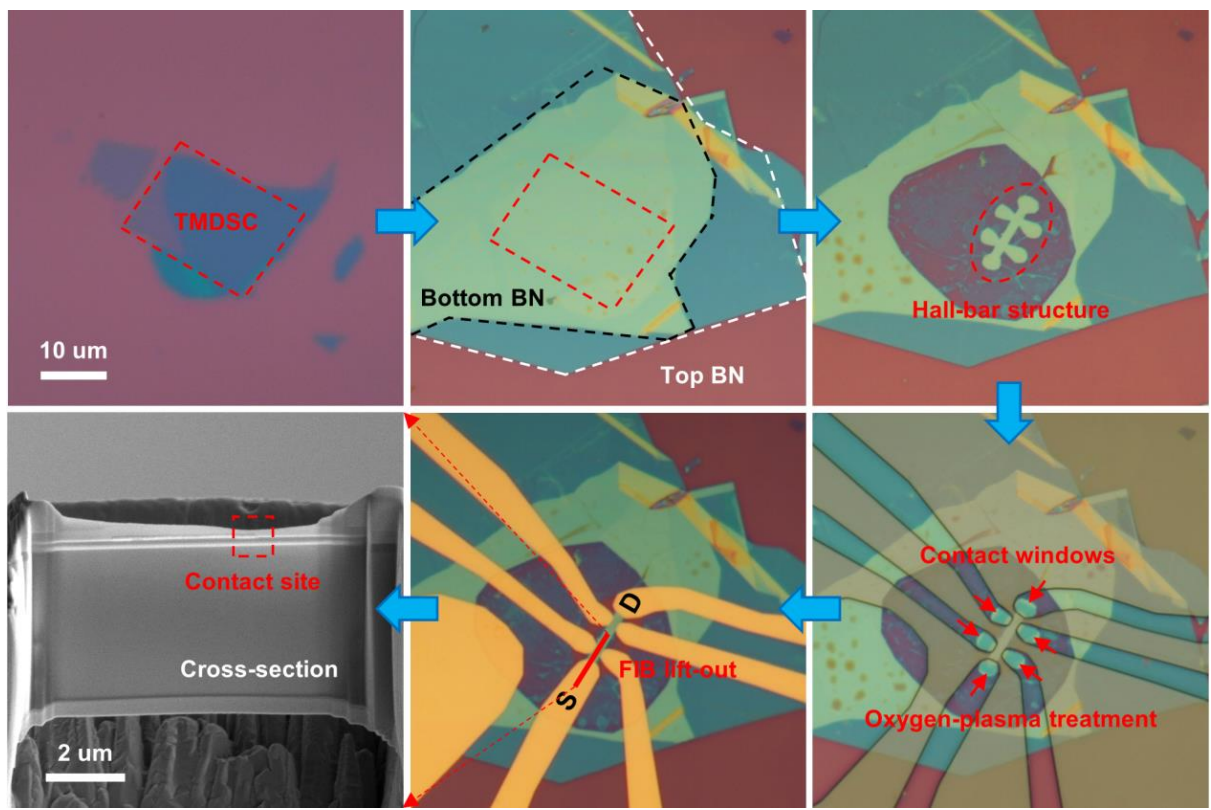

**Fig. S2: Sample fabrication steps.** Optical images showing the step-by-step FET fabrication by the LBD method: mechanical exfoliation; BN encapsulation by the dry transfer technique; Hall-bar shaping; contact windows by the referenced RIE processes; soft oxygen-plasma treatment; metal lead deposition by the electron-beam evaporation. After electrical measurements, cross-section samples from one of the main carrier-injection contacts were prepared by FIB techniques as shown in the SEM view.

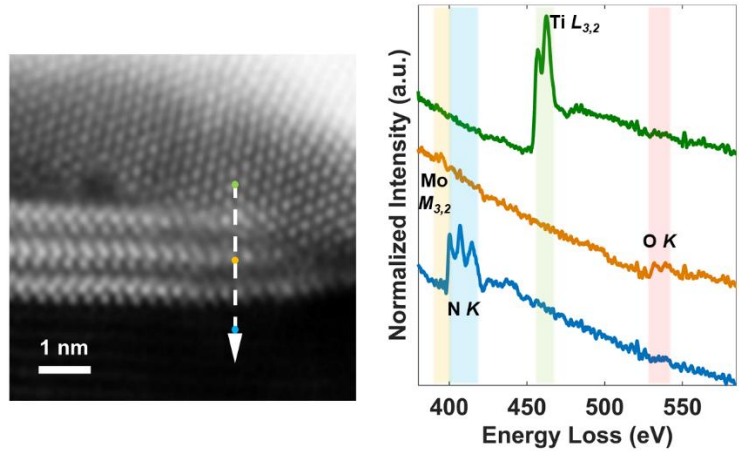

**Fig. S3: EELS analysis of the LBD contact region.** An EELS line-scan was performed across the metal/LBD/BN heterostructure as marked in the left ADF image. After aligning the energy-loss to the absolute energy scale according to the simultaneously recorded zero-loss peaks, the normalized spectra showing characteristic core-loss edges were plotted along the scan direction in the right panel. The oxygen K edge can be identified from the power-law background in the LBD region, suggesting the oxygen substitution.

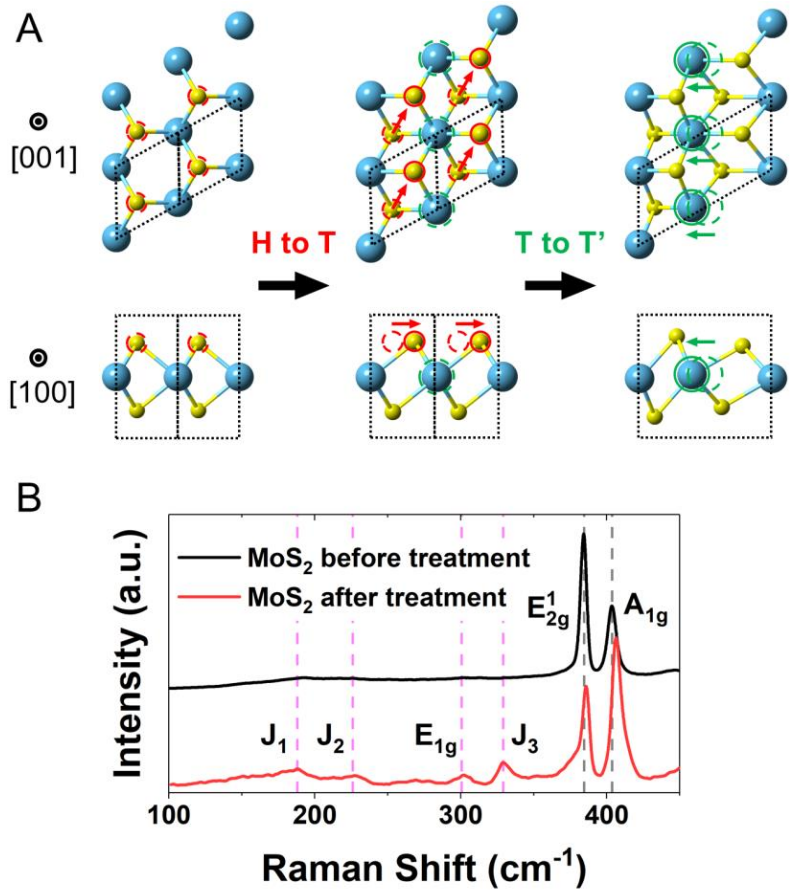

**Fig. S4: Raman spectroscopy of the LBD. (A)** Atomic structure transformation by TMDSC bond rearrangements (TM stands for transition metals and cyan balls in models; C stands for chalcogen atoms and yellow balls in models). Theoretically, only a slight sliding of top-layer chalcogen atoms along the <110> direction as shown by red arrows is necessary for the coordination change of [TM-C$_6$] polyhedra from the trigonal-prismatic coordination (H) to the octahedral one (T). Then a small shift of metal atoms along the <120> direction as shown by green arrows will lead to the distorted T structure (T'). Black parallelograms illustrate the unit cells. **(B)** Raman measurements on MoS$_2$ edges before and after the soft oxygen-plasma treatment. The blue shift and broadening of characteristic Raman modes, i.e. E$_{2g}$ (in-plane vibration) and A$_{1g}$ (out-of-plane vibration), indicate the occurrence of bonding distortion, which is consistent with the literature about Raman spectroscopy of lattice distortion (J. Raman Spectrosc. 2019, 50:1661-1671). There are also some emergent soft modes in the Raman spectrum of treated $MoS_2$ being assigned to the $J_1$, $J_2$ and $J_3$ Raman modes of 1T and 1T' $MoS_2$ structures as marked by magenta dashed lines, which originate from the octahedral coordination after the LBD process (Nano Lett. 2015, 15, 1, 346-353).

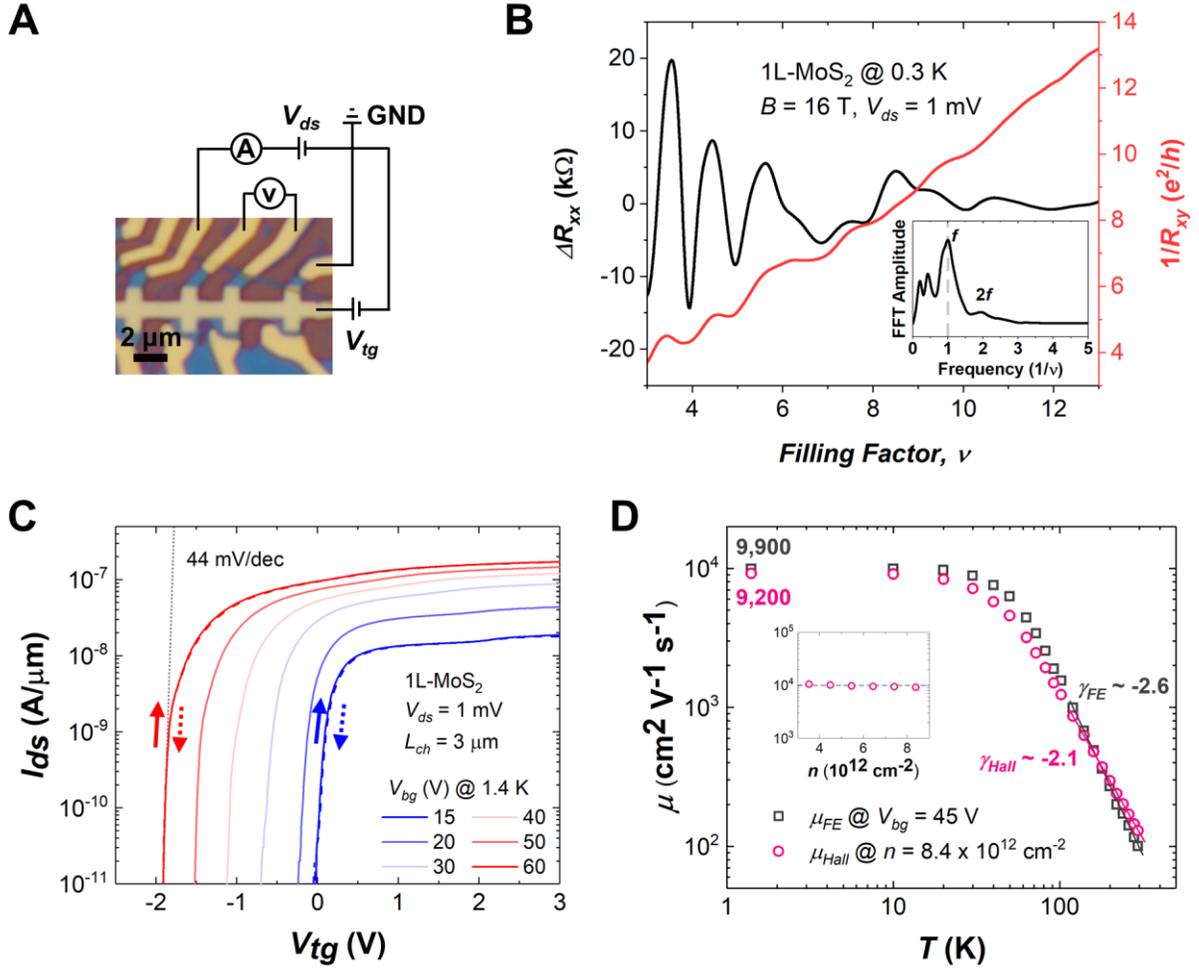

**Fig. S5: 1L-MoS₂ FET by LBD contacts.** (**A**) Optical image of a dual-gated 1L-MoS₂ FET utilizing the LBD contact method. (**B**) Quantum transport measurements showing prominent $R_{xx}$ oscillation and $R_{xy}$ quantization. The inset is the fast Fourier transform (FFT) amplitude of $R_{xx}$ versus filling factors, indicating the principal oscillation period of $\Delta\nu$=1. (**C**) Top-gated transfer curves with $V_{bg}$ varied from 15 to 60 V. Negligible hysteresis is shown for both upper and lower $V_{bg}$ limits. (**D**) Temperature-dependent characteristics of the field-effect and Hall mobilities. The Hall mobility increases slightly with lowered carrier concentrations at 1.4 K as the inset shows.

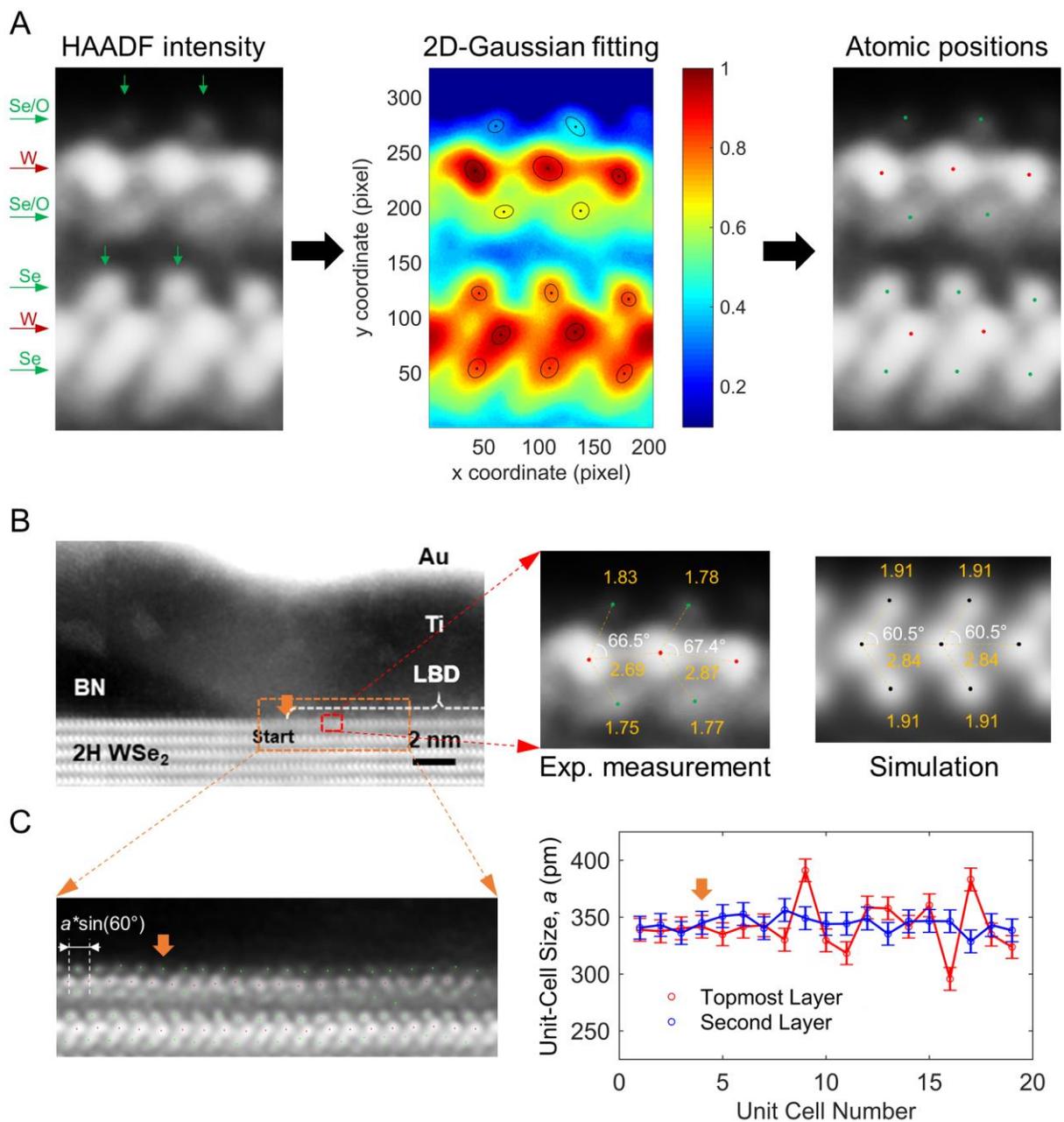

**Fig. S6: ACSTEM image analysis. (A)** The data-processing procedures to determine atomic positions, where red and green dots are the fitted W and Se atomic positions, respectively. The atomic positions were precisely determined by the 2D-gaussian fitting of normalized HAADF intensities of each atomic column. **(B)** Quantitative analysis of the structure distortion by bond lengths and angles. The WSe₂ layer at the contact region distorted locally by the shrinking of first

W-W bond length and the elongating of following one, together with shortened W-Se/O bond lengths and stretched Se-W-Se angles, when compared to the simulated standard structure (equal W-W, W-Se spacings and Se-W-Se angles in all unit cells). **(C)** The variation of distortion across a wider range and in different layers of $WSe_2$ is evaluated by calculating the lattice parameter, $a$, of each unit cell. It can be seen that the distortion starts right inside the contact region, as marked by orange arrows, showing locally shrunk and elongated unit-cell sizes. But the second layer is protected by the topmost layer and its unit-cell sizes remain relatively constant as that in the standard structure.

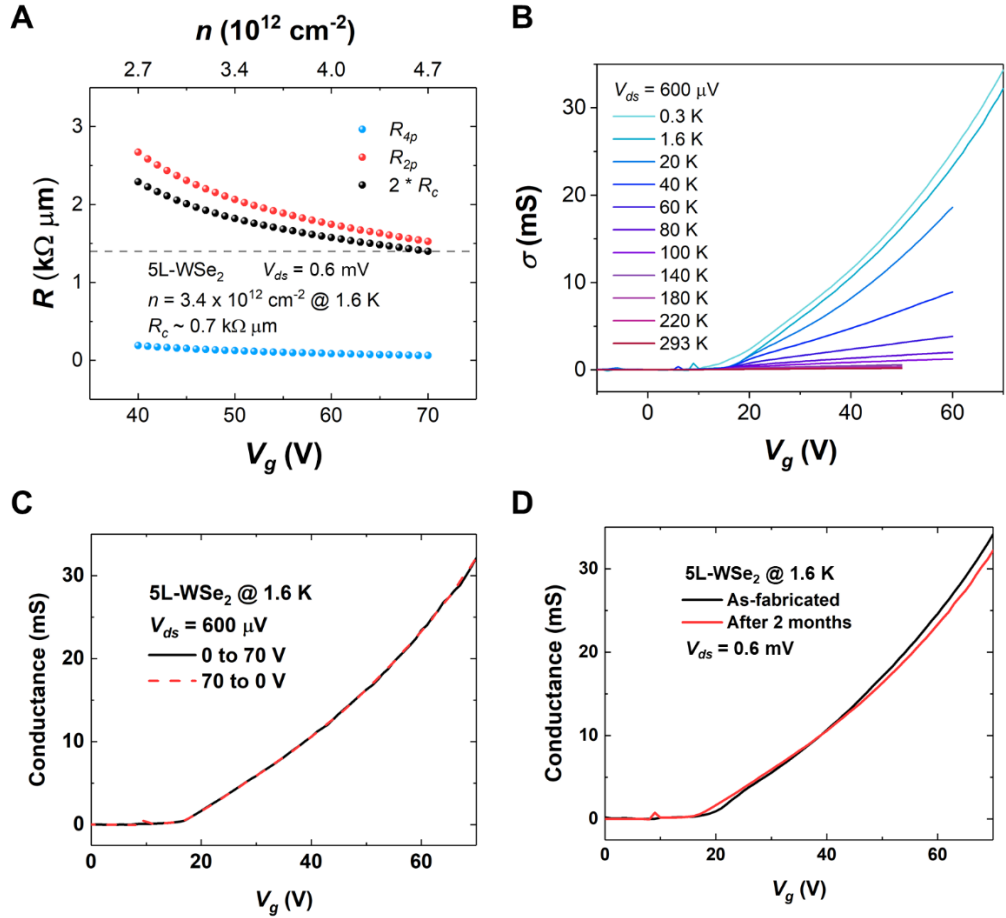

**Fig. S7: Extended data of the 5L-WSe₂ FET using LBD contacts.** **(A)** Extraction of contact resistance by four-probe measurements. **(B)** Temperature-varied transfer curves, from which field-effect mobilities at each temperature can be calculated. **(C)** Forward- and backward-scans of the channel conductance. **(D)** Transfer curves of the same 5L-WSe₂ FET right after the fabrication and after 2-month storage. The small shift of the threshold voltage may be due to the annealing effects when removing the sample from the cryogenic station after the first-round test.

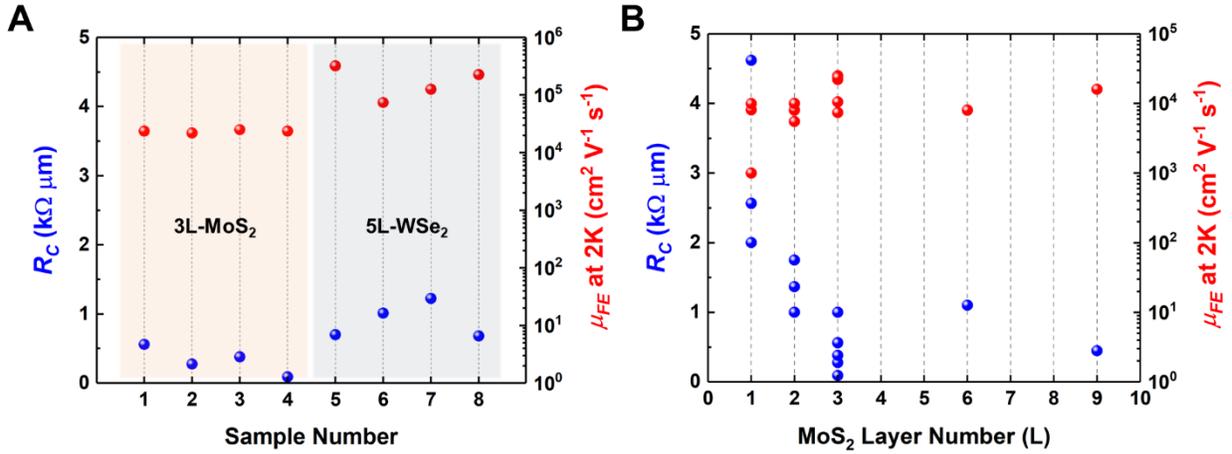

**Fig. S8: Reproducibility and practicality of LBD contacts in TMDSCs.** **(A)** Contact resistance and field-effect mobility of multiple back-gated 3L-MoS$_2$ and 5L-WSe$_2$ FETs using the LBD contact strategy. The high-quality contact is reproducible once suitable conditions are set up. The device deviation is reasonable considering the fabrication error by human handling and laboratory facilities. We also notice that WSe$_2$ FETs show a larger device-to-device variation, which may be attributed to the higher mixed state of octahedral derivatives than the situation of MoS$_2$. **(B)** Device performance of multiple MoS$_2$ FETs of different layer numbers. Both the achievable charge carrier mobility and contact resistance depend strongly on the channel thickness, i.e. thinner channel leads to suppressed mobility and higher contact resistance. The dramatically increased contact resistance and lowered carrier mobility for monolayers may be due to the weaker screening effect and thus increased scattering by impurity in the atomically thin channels.

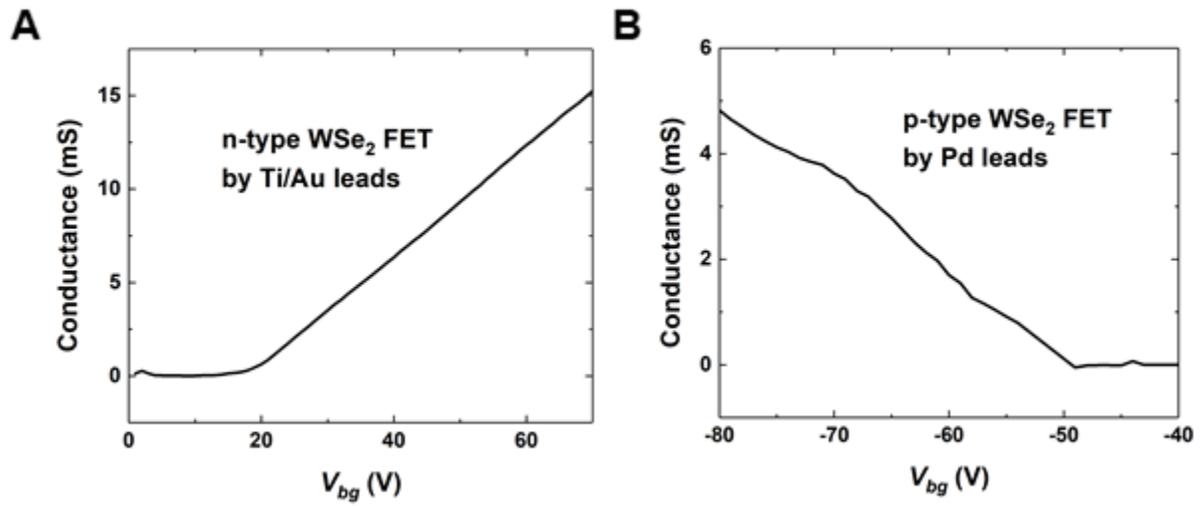

**Fig. S9: Lead-dependent polarity of WSe₂ FETs.** Low-temperature transfer curves of few-layer WSe₂ FETs fabricated by the same LBD contact method but with **(A)** Ti/Au and **(B)** Pd leads, respectively.

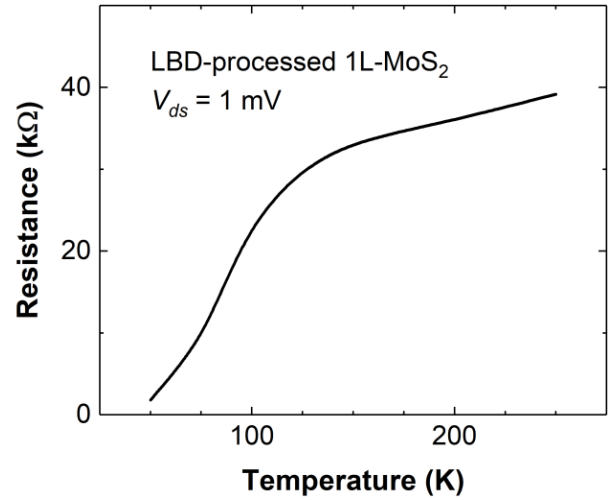

**Fig. S10: LBD-induced metallicity in TMDSC monolayers.** The resistance of 1L-MoS$_2$ after treated by the same soft oxygen-plasma process as fabricating LBD contacts was measured at varied temperatures. It can be seen that the resistance of processed 1L-MoS$_2$ shows the metallic behavior by continuously decreasing when the temperature lowers. The sharper drop of resistance below 150 K may originate from the de-saturation of those dispersed DOS near the Fermi level, suggesting semi-metallic characteristics.